\documentclass[11pt,a4paper]{article}
\usepackage{float}
\usepackage[T1]{fontenc}
\usepackage[utf8]{inputenc}
\usepackage{lmodern}
\usepackage{amsmath,amssymb,amsfonts,mathtools}
\usepackage{graphicx,booktabs,array,microtype,xcolor}
\usepackage[margin=2.45cm]{geometry}
\usepackage[colorlinks=true,linkcolor=blue!55!black,citecolor=blue!55!black,urlcolor=blue!55!black]{hyperref}
\usepackage[numbers,sort&compress]{natbib}
\newcommand{\scatt}{a_{\rm sc}}
\newcommand{\acrit}{a_{\rm crit}}
\newcommand{\mc}{\mathcal M}

\providecommand{\doi}[1]{\href{https://doi.org/#1}{\nolinkurl{doi:#1}}}
\newcommand{\arxivref}[1]{\href{https://arxiv.org/abs/#1}{\nolinkurl{arXiv:#1}}}

\definecolor{orcidlogocol}{HTML}{A6CE39}
\providecommand{\orcidlink}[1]{%
  \href{https://orcid.org/#1}{\hspace{1.5pt}\begin{picture}(8,8)(-1,-1)%
  \put(3,3){\color{orcidlogocol}\circle*{7.5}}%
  \put(1,1.2){\color{white}\fontfamily{phv}\fontsize{3.5}{3.5}\selectfont\bfseries iD}%
  \end{picture}\hspace{1pt}}%
}

\title{Scalar and pseudoscalar glueballonia: do they bind?}
\author{Narmin Nasibova\orcidlink{0000-0001-6475-6251}$^{1,2,3}$ and Francesco Giacosa\orcidlink{0000-0002-7290-9366}$^{4,5}$\\[0.5em]
\small $^1$Institute of Physics, Ministry of Science and Education, Baku, Azerbaijan\\
\small $^2$Center for Theoretical Physics, Khazar University, Baku, Azerbaijan\\
\small $^3$Baku Engineering University, Khirdalan city, Azerbaijan\\
\small $^4$Institute of Physics, Jan Kochanowski University, Kielce, Poland\\
\small $^5$Institute for Theoretical Physics, J.W. Goethe University, Frankfurt am Main, Germany}
\date{}

\begin{document}
\maketitle
\begin{abstract}
We investigate the existence of molecular two-glueball bound states,
so-called \emph{glueballonia}, in pure Yang--Mills (YM) theory.
We mainly concentrate on the scalar and pseudoscalar channels, thus studying
scalar--scalar and scalar--pseudoscalar bound states, respectively.
Near the onset of binding, when a bound state first appears just below
threshold, a nonrelativistic framework is justified.
We employ an attractive scalar-exchange Yukawa potential together with
a smeared contact term contribution and determine the critical attraction
required for binding.
For two scalar glueballs, we use the scalar dilaton effective theory to determine the parameters of the non-relativistic potential: the Yukawa interaction is generated by t- and u-channels involving the exchange of one scalar glueball, while the repulsion emerges by the s-channel annihilation and contact interaction.  Using lattice-YM input to fix the dilaton parameters, we find that the attraction is strong enough: the scalar glueballonium binds, and the result is quite robust with respect to parameter variations. In a minimal scalar--pseudoscalar extension, the Yukawa attraction is again given by the t-channel scalar glueball, while the crossed u-channel pseudoscalar exchange induces an attractive local contribution, making the pseudoscalar glueballonium also bind in pure gauge. Importantly, both the scalar and pseudoscalar glueballonia are stable in the YM vacuum. Next, we briefly comment on the pseudoscalar-pseudoscalar, scalar-tensor, and tensor-tensor channels: all these channels are attractive too and could form bound states, which, however, would be unstable in pure YM due to decay into two lighter glueballs. Finally, as a cross-check, our formalism confirms that Higgs-Higgs bound state does not form.
\end{abstract}

\section{Introduction}\label{sec:intro}

Glueballs are color-singlet excitations built entirely from the gluonic degrees of freedom of quantum chromodynamics (QCD). The existence of such states is a direct consequence of non-Abelian gauge dynamics; their identification in experiment is complicated by mixing with conventional mesons, nearby thresholds, and hadronic decays; for recent reviews about glueball candidates, see Refs.~\cite{Mathieu:2008me,Ochs:2013gi,Llanes-Estrada:2021evz,Chevalier:2026Glueballs}.

In the pure Yang--Mills (YM) limit, quark degrees of freedom and decay channels into conventional quark-antiquark mesons are absent. Lattice YM calculations then predict a discrete low-lying glueball spectrum, with the lightest scalar followed by tensor and pseudoscalar excitations, \cite{Morningstar:1999rf,meyer2005glueball, Chen:2005mg,Athenodorou:2020ani,Morningstar:2025Update}, see also their mutual comparison in Ref. \cite{Trotti:2022knd}. These low-lying glueballs are stable in pure YM. 
This fact makes glueball--glueball interactions, and especially possible bound states of two glueballs, a well-defined subject of investigation.

The possibility that two scalar glueballs form a composite bound state, denoted as \textit{glueballonium}, was first studied in Ref.~\cite{Giacosa:2021brl}. Within a dilaton effective Lagrangian, relativistically unitarized $GG$ scattering generates a bound-state pole once the attraction becomes sufficiently strong. For lattice-motivated parameters, a scalar glueballonium close to threshold was predicted. However, the quantitative conclusions retain some dependence on the unitarization procedure, motivating a complementary analysis.

The present work returns to the formation point from a complementary direction: rather than determining an entire relativistic scattering amplitude, we ask when the first two-body level appears in a nonrelativistic (NR) potential, for which the zero-energy Schr\"odinger equation provides a well-defined non-perturbative criterion.
In fact, for constituent masses $m_1,m_2$, reduced mass $\mu=m_{1}m_{2}/(m_1+m_2)$, and binding energy $B$, the binding momentum $\kappa=\sqrt{2\mu B}$ tends to zero as $B\to 0$ \cite{guo2018hadronic, hanhart2025hadronic}. Here the long-distance motion is then nonrelativistic, and the formation criterion may be studied with familiar potential-model methods \cite{Lucha:1989PotentialModels,Lucha:1991vn, Braaten:2004rn}. 

At the same time, deriving a nonrelativistic interaction from a relativistic field theory requires matching, here naturally provided by the dilaton/scalar glueball Lagrangian as an effective low-energy YM theory \cite{Migdal:1982jp,Salomone:1980sp}. Namely, exchange of the lightest scalar glueball in the t- and u-channels between two glueballs generates an attractive Yukawa interaction in the NR limit \cite{Yukawa:1935xg}. The dilaton potential also contains a direct four-glueball vertex that, together with the s-channel annihilation channel, generates an additional contact term repulsion. This term requires special care: an ordinary Dirac delta is unproblematic as a potential in one spatial dimension, but in general a three-dimensional point interaction needs a regularization, renormalization, or self-adjoint-extension prescription \cite{Beg:1985NR,Jackiw:1991Delta,Gosdzinsky:1991Delta,Geltman:2011Delta,Farrell:2010ZeroRange}. A finite-length regularization is suitable for the present case, since glueballs are composite objects with a finite radius: the contact term is smeared out. We therefore calculate the binding threshold as a function of both the Yukawa attraction and the strength and spatial range of the contact term. Matching the interaction to the dilaton couplings at the central lattice-motivated parameters yields (at least) one $GG$ bound state: the scalar glueballonium is predicted to exist. Moreover, since it is lighter than the lowest two-glueball threshold, it is a stable state in pure YM theory. 
This conclusion is robust under variation of the parameters of the model.

%neither a small binding energy nor the Schr\"odinger equation itself fixes the smeared contact term coefficients \cite{Tong:2006QFT,Padmanabhan:2018NR}. This matching is  
%This distinction is central to our analysis. We do not claim that a near-threshold treatment predicts the mass of a deeply bound glueballonium with controlled accuracy.

We extend the discussion beyond the lightest scalar pair. In particular, a mixed scalar--pseudoscalar $GP$ pair has $J^{PC}=0^{-+}$ in an $S$ wave.  The t-channel involves a scalar glueball exchange and generates the Yukawa attraction. The u-channel crossed interaction contains a pseudoscalar glueball exchange that, within the local-threshold approximation, reduces to an attractive contact interaction that is larger in absolute value than the s-channel and four-leg interaction's contributions: the resulting contact interaction is here attractive. We find that the $GP$ channel also binds over the parameter range investigated; importantly, the Yukawa attraction is sufficiently strong to bind even without the additional attractive contact term. In this respect, the existence of a bound state does not depend on the detail of the short-range term. This $0^{-+}$ molecule, called \textit{pseudoscalar glueballonium}, is also stable against strong decay in pure YM for a CP-parity-conserving YM theory (hence, without the $\theta$-term).

We comment also on pseudoscalar-pseudoscalar, scalar-tensor, and tensor-tensor glueballonia. Although attraction, and possibly bound states, also occurs in these channels, the corresponding putative states are not stable in pure YM, since they can decay into lighter glueball pairs. They might rather appear as resonances in pure YM theory, making their identification more challenging on the lattice. 
Finally, as a cross-check, we apply our formalism also to Higgs-Higgs interaction, recovering a known result: no Higgsonium forms in the Standard Model. 
Glueball interactions and composite configurations
have attracted attention from different theoretical perspectives.
Lattice determinations of inter-glueball potentials, particularly in $SU(2)$ Yang--Mills theory, have found significant attraction and provided low-energy scattering information \cite{deForcrand:1985,Yamanaka:2020qom,Yamanaka:2021xqh,Yamanaka:2019pot,Chu:2024Glueball,Morningstar:2025Update}. These studies support a potential-based approach, although their differing Yukawa fits and gauge groups prevent a direct matching to the $SU(3)$ dilaton model. In addition, recent lattice calculations of the scalar glueball gravitational form factors indicate a relatively small mass radius, $R_G=0.263(31)$ fm \cite{abbott2026lattice}. This supplies a useful physical length scale for estimating the spatial range of a local four-glueball interaction. 
%The dependence on the regulator radius $r_0$ is therefore an important part of our investigation.
In related phenomenological approaches, glueball-meson molecules as possible components
of excited hadrons were studied in Ref. \cite{Petrov:2022ipv}, whereas glueballonia as topological Hopfions, predicting both
tightly and loosely bound configurations, are described in Ref. \cite{Amari:2025cre}.
In a complementary work, S-matrix bootstrap methods have
been used to constrain glueball couplings \cite{Guerrieri:2023qbg},
while a recent effective field theory study examined scalar-glueball
form factors and interaction radii \cite{Issifu:2026tjd}.
Within this topic, our work aims to show that novel states in pure YM gauge theory are expected to exist, motivating further research in this direction. 
The paper is organized as follows. Section~\ref{sec:potential} introduces the Yukawa-plus-contact problem and its $S$-wave threshold condition. Section~\ref{sec:applications} applies the construction to $GG$, $GP$, and exploratory $PP$, $GT$, and $TT$ pairs and compares it with lattice information; a short excursus on Higgs-Higgs is also presented.  Section~\ref{sec:conclusions} presents a brief summary and outlook. The appendix~\ref{app:app} discusses the radius scan, compares alternative finite-range realizations, and  presents the pseudoscalar glueballonium beyond the local approximation. Unless stated otherwise, $G$, $P$, and $T$ denote the $0^{++}$, $0^{-+}$, and $2^{++}$ glueballs in parity-conserving pure Yang--Mills theory, with $\hbar=c=1$.

\section{Yukawa interaction and a smeared Dirac delta}\label{sec:potential}

\subsection{Potential, scaling, and binding criterion}\label{sec:general}

For two particles with reduced mass $\mu=m_1m_2/(m_1+m_2)$, we study the NR potential
\begin{equation}\label{eq:potential}
 V(r)=-\alpha\frac{e^{-mr}}{r}+\beta\,\delta_{r_0}^{(3)}(\boldsymbol r)\text{ .}
 \qquad \alpha\geq0,\quad \beta\in\mathbb R \text{ .}
\end{equation}
The exchanged mass is $m$, the Yukawa coupling is $\alpha$, the contact coefficient is $\beta$ with units of mass$^{-2}$, and the normalized spherical regulator is taken as
\begin{equation}\label{eq:delta}
 \delta_{r_0}^{(3)}(\boldsymbol r)=\frac{3}{4\pi r_0^3}\Theta(r_0-r) \text{ , } \int d^3r\delta_{r_0}^{(3)}(\boldsymbol r)=1 \text{ . }
\end{equation}
Thus, $\beta>0$ is repulsive and $\beta<0$ attractive. The $S$-wave radial equation is
\begin{equation}\label{eq:radial}
 -\frac{1}{2\mu}\frac{d^2u}{dr^2}+V(r)u(r)=Eu(r) \text{ , }
  u(0)=0 \text{ , } \psi(\boldsymbol r)=\frac{u(r)}{\sqrt{4\pi}\,r} \text{ . }
\end{equation}
At zero energy, the outer solution ($r \rightarrow \infty$) is asymptotically linear:
\begin{equation}\label{eq:asymptotic}
 u(r)=A+Cr+o(1) \text{ , } \scatt=-\frac{A}{C} \text{ ,}
\end{equation}
where $\scatt$ is the scattering length. 
The \emph{first} binding threshold occurs for $C=0$: the regular, nodeless radial solution approaches a constant and the scattering length diverges. This is the standard zero-energy resonance criterion; the sign and magnitude of the scattering length govern the nearby bound or virtual pole \cite{Braaten:2004rn,Farrell:2010ZeroRange}. This marginal zero-energy state is not normalizable. Below threshold, a bound solution has $E=-B<0$ and $u(r)\propto\exp(-\sqrt{2\mu B}\,r)$.

Introducing the \emph{dimensionless} quantities
\begin{equation}\label{eq:dimless}
 x=mr,\quad x_0=mr_0,\quad a=\frac{2\mu\alpha}{m},\quad
 b=2\mu m\beta,\quad \epsilon=\frac{2\mu E}{m^2},
\end{equation}
Eq. (\ref{eq:radial}) becomes:
\begin{equation}\label{eq:scaledradial}
 \left[-\frac{d^2}{dx^2}-a\frac{e^{-x}}{x}
 +\frac{3b}{4\pi x_0^3}\Theta(x_0-x)\right]u(x)=\epsilon u(x).
\end{equation}
For this regulator, the first $S$-wave threshold is a surface $\acrit\equiv\acrit (b,x_0)$. The pure-Yukawa result reads \cite{Edwards:2017Yukawa}:
\begin{equation}\label{eq:yukawa}
 \acrit(b=0,x_0)=a_Y=1.6798078\ldots,\qquad
 \alpha_{\rm crit}^{Y}=\frac{m}{2\mu}a_Y \text{ ;}
\end{equation}
see also the spectral estimates, numerical calculations, and systematic approximations of Refs.~\cite{Hall:1992Yukawa,Li:2006Yukawa,Napsuciale:2021Yukawa}.

Next, we determine the onset of bound states by solving
Eq.~\eqref{eq:scaledradial} at $\epsilon=0$, with the
regular origin expansion, valid for $0<x\ll x_0$,
\begin{equation}\label{eq:originseries}
 u(x)=x-\frac{a}{2}x^2
 +\frac{a^2/2+a+K}{6}x^3+O(x^4)
 \text{ , } K=\frac{3b}{4\pi x_0^3} \text{ . }
\end{equation}
A continuous phase $u=R\sin\vartheta$,
$u'=R\cos\vartheta$ is convenient for identifying
the first threshold:
\begin{equation}\label{eq:phase}
 \vartheta'(x)=\cos^2\vartheta(x)
 -\left[-a\frac{e^{-x}}{x}
 +K\Theta(x_0-x)\right]\sin^2\vartheta(x) \text{ , }
  \vartheta(0)=0 \text{ . }
\end{equation}
The threshold is determined by requiring
$u'(x)\to0$ as $x\to\infty$, with a nodeless
zero-energy solution. Numerically, this condition
is imposed at a sufficiently large finite radius.
Independent shooting of the radial equation and
variation of the large-distance integration cutoff
provide cross-checks. The results are displayed
in Fig.~\ref{fig:critical}, with a separate examination
of the small-radius limit in Sec. 3 and in Appendix~\ref{app:app}.

\begin{figure}[tb]
 \centering\includegraphics[width=0.82\linewidth]{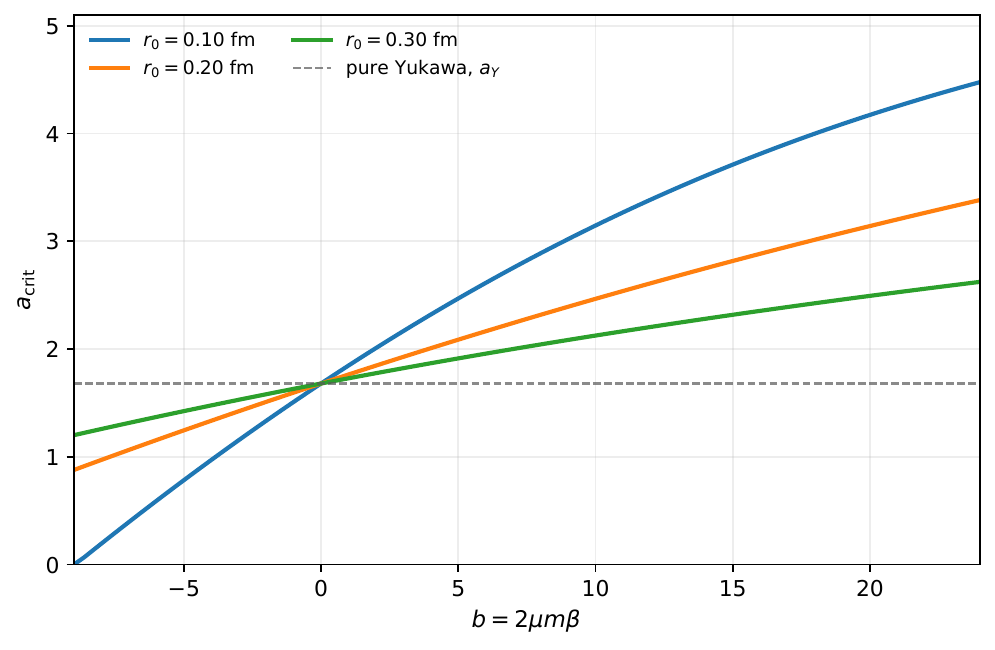}
 \caption{First $S$-wave threshold $\acrit(b,x_0)$ for three physical radii using $m=m_G=1.710$ GeV. The dotted horizontal reference  is the pure-Yukawa value $a_Y$. }
 \label{fig:critical}
\end{figure}

%Thus, the Schr\"odinger equation provides a nonperturbative determination of the binding threshold for a specifiedpotential, without requiring an  additional unitarization prescription. The remaining uncertainty lies in matching the potential to the underlying field theory, particularly in its smeared contact term component. 

\subsection{The contact-only solution}\label{sec:threshold}

The contact-only problem ($\alpha=0$) is exactly soluble and deserves a separate discussion because it helps to understand the meaning and the need of a nonzero parameter $r_0$.
For a \emph{repulsive} spherical barrier ($\beta>0$),
matching the zero-energy wavefunctions at $r_0$ gives
\begin{equation}\label{eq:repulsive_length}
 \displaystyle
 \scatt^{(+)}=r_0\left(1-\frac{\tanh z_+}{z_+}\right) \text{ , }
 z_+^2=\frac{3\mu\beta}{2\pi r_0}>0 \text{ . }
\end{equation}
Since $0\leq\scatt^{(+)}<r_0$, the scattering length
necessarily vanishes as $r_0\to0$, regardless of how
the nonnegative strength $\beta$ depends on $r_0$.
More generally, the entire scattering amplitude vanishes
at any fixed finite momentum. Thus, a repulsive interaction
becomes ineffective in the zero-range limit, even if its
integrated strength remains finite or diverges.
This behavior provides an example of nonrelativistic
triviality. Indeed, the finite range $r_0$ acts as an
ultraviolet regulator, with a characteristic momentum
scale $\Lambda\sim 1/r_0$. Removing this regulator
causes the physical scattering amplitude to vanish,
despite the presence of a repulsive bare interaction.
This is connected to the nonrelativistic
triviality of $\lambda\varphi^4$ theory established
in Ref. \cite{Beg:1985NR}; see also
Ref.~\cite{Gosdzinsky:1991Delta}.
A related phenomenon occurs in relativistic
$\lambda\varphi^4$ theory in four spacetime
dimensions, whose continuum limit is likewise
trivial \cite{Aizenman:2021Triviality}.
The underlying ultraviolet behavior is, however,
different: the relevant divergence is linear in
the cutoff in the present nonrelativistic problem,
whereas it is logarithmic in the relativistic theory.

On the other hand, for an \emph{attractive} core ($\beta<0$), the trigonometric expression instead is
\begin{equation}\label{eq:attractive_length}
\displaystyle
 \scatt^{(-)}=r_0\left(1-\frac{\tan z_-}{z_-}\right),\qquad
 z_-^2=\frac{3\mu|\beta|}{2\pi r_0}.
\end{equation}
The first contact-only bound state appears at $z_-=\pi/2$, or
\begin{equation}\label{eq:contactcrit}
 \beta_{\delta,\rm crit}=-\frac{\pi^3r_0}{6\mu}<0 \text{ ,}\qquad
 b_{\delta,\rm crit}=-\frac{\pi^3x_0}{3}<0 \text{ .}
\end{equation}
Thus, for $b<b_{\delta,\rm crit}$, the attraction
provided by the core already binds at $a=0$; our
threshold plots therefore set the \emph{minimal
nonnegative} $\acrit$ to zero.

Equations~\eqref{eq:repulsive_length}--\eqref{eq:contactcrit}
also reveal a fundamental difference between
repulsive and attractive contact interactions
in three dimensions. In the repulsive case,
the scattering length satisfies
$0\leq a_{\rm sc}<r_0$ and necessarily vanishes
as $r_0\to0$, irrespective of how the barrier
strength is varied. An attractive core, instead,
can admit a nontrivial zero-range limit if its
strength is appropriately tuned to keep a finite
scattering length. This is possible both for
$a_{\rm sc}>0$, corresponding to a bound state,
and for $a_{\rm sc}<0$, corresponding to an
attractive interaction without a bound state
(on the first resonance branch).

For the reasons described above, throughout this work
we retain a finite interaction radius $r_0$, both for
the repulsive contact interaction in the $GG$ channel,
which becomes trivial in the zero-range limit, and
for the effective attractive contact in the $GP$
channel, where a nontrivial zero-range limit would
require an appropriate tuning of the coupling.
It is worth noting that the direct four-point
interaction in the $GP$ channel is actually repulsive:
the net attraction of the contact term arises from
the local approximation to crossed pseudoscalar
exchange. When the latter is retained explicitly,
the remaining contact interaction is again repulsive,
as discussed later on and in Appendix \ref{app:app}.
Thus, we regard $r_0$ as a physical model parameter
associated with the finite spatial extent of
the interaction.

\subsection{From scattering amplitudes to potentials}
\label{sec:QFTmatching}

\begin{figure}[tb]
\centering
\includegraphics[width=0.83\linewidth]{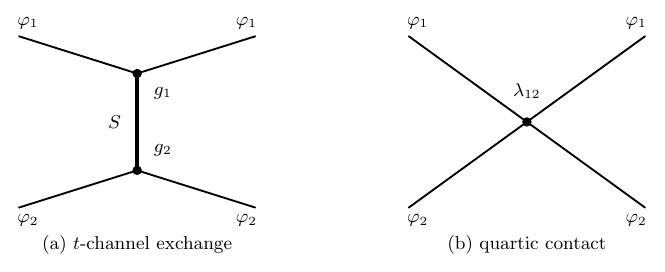}
\caption{Tree-level $\varphi_1\varphi_2\to\varphi_1\varphi_2$
scattering: (a) $t$-channel exchange of the scalar $S$
(thicker internal line), with vertices $g_1$ and $g_2$;
(b) the direct quartic vertex $\lambda_{12}$. All 
lines represent scalar fields.}
\label{fig:generic_matching}
\end{figure}

We briefly recall how to obtain a nonrelativistic potential
from a relativistic scattering amplitude calculated using
standard Feynman rules
\cite{Peskin:1995ev}.
Consider two distinguishable scalar fields $\varphi_1$ and
$\varphi_2$, with masses $m_1$ and $m_2$, interacting by
exchange of a scalar field $S$ of mass $m_S$. We also allow
for a direct quartic interaction,
\begin{equation}
\mathcal L_{\mathrm{int}}
=-\frac{g_1}{2!}S\varphi_1^2
 -\frac{g_2}{2!}S\varphi_2^2
 -\frac{\lambda_{12}}{2!\,2!}\varphi_1^2\varphi_2^2 \text{ . }
\label{eq:general_interaction}
\end{equation}
The corresponding vertex factors are $-ig_1$, $-ig_2$,
and $-i\lambda_{12}$. The two tree-level contributions to
$\varphi_1\varphi_2\to\varphi_1\varphi_2$ are illustrated
in Fig.~\ref{fig:generic_matching}.
With the usual Feynman diagram convention,
we obtain at tree-level t-channel and contact amplitudes as: 
\begin{equation}
\begin{aligned}
i\mathcal M_t
=(-ig_1)\frac{i}{t-m_S^2+i0}(-ig_2)\text{  ,  }
i\mathcal M_4&=-i\lambda_{12} \text{ ,}
\end{aligned}
\end{equation}
leading to the total amplitude as
\begin{equation}
\mathcal M
=-\frac{g_1g_2}{t-m_S^2}-\lambda_{12} \text{ .}
\label{eq:general_tree_amplitude}
\end{equation}
In the nonrelativistic limit the energy transfer is
negligible. With $\mathbf q=\mathbf p'-\mathbf p$,
we have $t\simeq-\mathbf q^2$ and thus
\begin{equation}
\mathcal M_{\mathrm{NR}} \equiv \mathcal M_{\mathrm{NR}}(\mathbf{q})
=\frac{g_1g_2}{\mathbf q^2+m_S^2}-\lambda_{12} \text{ .}
\label{eq:general_NR_amplitude}
\end{equation}

To relate this amplitude to a potential, recall that
for elastic scattering in the center-of-mass frame,
standard relativistic state normalization gives, near
threshold,
\begin{equation}
\frac{d\sigma}{d\Omega}
=\frac{|\mathcal M_{\mathrm{NR}}|^2}
 {64\pi^2(m_1+m_2)^2}=|f_{\mathrm B}|^2 \text{ , } f_{\mathrm B}= \frac{\mathcal M_{\mathrm{NR}}}{8\pi(m_1+m_2)} 
\end{equation}
where in the r.h.s. the usual phase convention has been adopted.
On the other hand, the first Born approximation in
quantum mechanics reads
\begin{equation}
f_{\mathrm B}(\mathbf q)
=-\frac{\mu}{2\pi}\int d^3r\,
 e^{-i\mathbf q\cdot\mathbf r} V(r) \text{ , }
\mu=\frac{m_1m_2}{m_1+m_2} \text{ .}
\end{equation}
Equating the two expressions and taking the inverse
Fourier transform leads directly to
\begin{equation}
\displaystyle
V(r)=-\frac{1}{4m_1m_2}
\int\!\frac{d^3q}{(2\pi)^3}\,
 e^{i\mathbf q\cdot\mathbf r}
\mathcal M_{\mathrm{NR}}(\mathbf q) \text{ .}
\label{eq:QFT_potential_matching}
\end{equation}
In particular, the factor $1/(4m_1m_2)$ follows from
the conversion between relativistically normalized
scattering states and nonrelativistic two-body states.

Substituting Eq.~\eqref{eq:general_NR_amplitude} into Eq.~\eqref{eq:QFT_potential_matching} and
using the standard Fourier transforms gives
\begin{equation}
\displaystyle
V(r)=-\frac{g_1g_2}{16\pi m_1m_2}
\frac{e^{-m_Sr}}{r}
+\frac{\lambda_{12}}{4m_1m_2}
\delta^{(3)}(\mathbf r) \text{ .}
\label{eq:general_QFT_potential}
\end{equation}
We therefore recover precisely the Yukawa-plus-contact
structure of Eq.~\eqref{eq:potential}, with
\begin{equation}
\alpha=\frac{g_1g_2}{16\pi m_1m_2} \text{ , }
\beta=\frac{\lambda_{12}}{4m_1m_2}\text{ . }
\label{eq:general_alpha_beta}
\end{equation}
For $g_1g_2>0$, scalar exchange is attractive; a positive
$\lambda_{12}$ gives a repulsive short-range interaction.

It is useful to recall when the Yukawa attraction alone
can bind. Setting $\lambda_{12}=0$, the first $S$-wave
bound state appears if
\begin{equation}
\displaystyle
\alpha>\alpha_{\mathrm{crit}}^{Y}
=\frac{m_S}{2\mu}a_Y\text{ ,}
\qquad a_Y\simeq1.6798 \text{ ,}
\label{eq:general_Yukawa_threshold}
\end{equation}
or, equivalently,
$g_1g_2>8\pi a_Y m_S(m_1+m_2)$.
With a quartic interaction, the contact term must be
treated together with the Yukawa attraction. Once a finite
range $r_0$ is specified, its effect on the binding threshold
is described by the dimensionless function
$a_{\mathrm{crit}}\equiv a_{\mathrm{crit}}(b,x_0)$ of Sec.~\ref{sec:general},
illustrated in Fig.~\ref{fig:critical}.

%Finally, the appearance of a three-dimensional delta potential in Eq.~\eqref{eq:general_QFT_potential} requiressome care. The tree-level matching determines its coefficient, but a nonperturbative point interaction in three spatial dimensions is not defined by the naive delta alone \cite{Beg:1985NR,Gosdzinsky:1991Delta,Farrell:2010ZeroRange}. We therefore use the finite-range realization introduced in Eq.~\eqref{eq:delta}. Its zero-range limit is discussedin Appendix~\ref{app:app}; the dependence on its range and profile is examined in Appendices~\ref{app:app} and \ref{app:delta_realizations}. The additional factorsassociated with identical particles will be addressed in the $GG$ application below.

\section{Applications to glueball pairs}\label{sec:applications}

In this section we study the emergence of glueballonia in different channels, with particular attention to the scalar-scalar and scalar-pseudoscalar bound states. We use the central lattice glueball masses $m_G=1.710$ GeV,  as well as the pseudoscalar and tensor masses $m_P=2.560$ GeV, and $m_T=2.390$ GeV from Ref.~\cite{Chen:2005mg}.

\subsection{Scalar Glueballonium: \texorpdfstring{$GG$}{GG}}\label{sec:GG}

\subsubsection{From the Dilaton Potential to \texorpdfstring{$GG$}{GG} Interactions}

We employ the standard dilaton potential for a scalar glueball field, as motivated by the gluonic trace anomaly \cite{Migdal:1982jp,Giacosa:2021brl},
\begin{equation}\label{eq:dilaton}
 V_{\rm dil}(G)=\frac{\lambda_G}{4}G^4
 \left[\ln\left(\frac{G}{\Lambda_G}\right)-\frac14\right] \text{ .}
\end{equation}

The dilaton parameters are fixed using the lattice
determinations of the scalar glueball mass and its
gluonic matrix element by Chen \emph{et al.}
\cite{Chen:2005mg}. With their normalization, the
input values are
\begin{equation}\label{eq:GG_lattice_inputs}
m_G=(1.710\pm0.090)\ \mathrm{GeV}
\text{ , }
S_G\equiv
\langle 0|g^2\mathrm{Tr}\,G_{\mu\nu}G^{\mu\nu}|G\rangle
=(15.6\pm3.2)\ \mathrm{GeV}^3
\text{ .}
\end{equation}
Using the trace-anomaly matching relation
\begin{equation}
S_G=\frac{48\pi^2}{11N_c}\lambda_G\Lambda_G^3
=\frac{48\pi^2}{11N_c}m_G^2\Lambda_G
\text{ , }
\end{equation}
one obtains $\Lambda_G=(0.372\pm0.086)$ GeV
and $\lambda_G=21\pm11$
\cite{Giacosa:2026theta}.
These are model-matched parameters, and not
direct determinations of two-glueball scattering.
The pure-gauge treatment is compatible with large-$N_c$ arguments, but large-$N_c$ counting by itself does not determine the finite-$N_c$ values of the dilaton couplings or of the two-glueball scattering length \cite{tHooft:1973alw,Giacosa:2024scx}. 
We recall also that the dilaton potential is the starting point of the extended Linear Sigma Model for low-energy QCD  \cite{Giacosa:2024epf,Parganlija:2012fy} and the state $G$ can be interpreted as the resonance $f_0(1710)$ \cite{Janowski:2014ppa}.

After shifting the vacuum expectation value, $G\mapsto\Lambda_G+G$ (we retain the symbol $G$ for the fluctuation), one obtains
\begin{align}\label{eq:dilexpansion}
 V_{\rm dil}(\Lambda_G+G)
 &=V_{\rm dil}(\Lambda_G)+\frac{m_G^2}{2}G^2
 +\frac{g_3}{3!}G^3+\frac{g_4}{4!}G^4+O(G^5) \text{ ,}\\
 & m_G^2=\lambda_G\Lambda_G^2 \text{ ,} \qquad
 g_3=\frac{5m_G^2}{\Lambda_G}\text{ ,}\qquad
 g_4=\frac{11m_G^2}{\Lambda_G^2} \text{ .}
\end{align}

\begin{figure}[tb]
\centering
\includegraphics[width=\linewidth]{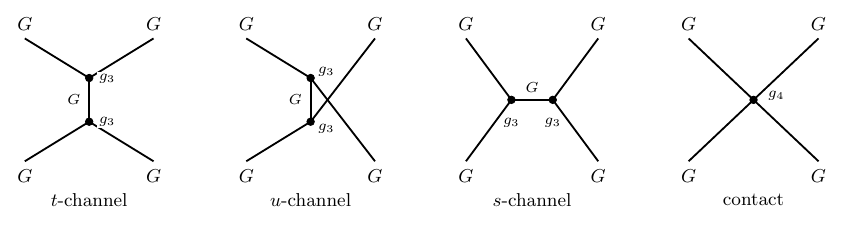}
\caption{All four contributions to the
$GG\to GG$ amplitude: scalar exchange in the
$t$, $u$, and $s$ channels, and the local
four-glueball vertex. External lines represent
identical scalar glueballs. }
\label{fig:GG_diagrams}
\end{figure}

The cubic and quartic interactions in
Eq.~\eqref{eq:dilexpansion} give rise to the four
diagrams in Fig.~\ref{fig:GG_diagrams}. 
For
$\mathcal L_{\rm int}=-g_3G^3/3!-g_4G^4/4!$,
the tree-level invariant amplitude reads
\begin{equation}\label{eq:GGamplitude}
\mathcal M_{GG}
=
-g_3^2\left[
\frac{1}{t-m_G^2}
+\frac{1}{u-m_G^2} + \frac{1}{s-m_G^2}
\right] -g_4 .
\end{equation}

For identical scalar glueballs, the nonrelativistic
Born amplitude must be Bose symmetrized. In particular,
the $t$- and $u$-channel diagrams correspond to the
direct and exchange contributions of a single Yukawa
potential, rather than to two independent interactions.
The local four-point and $s$-channel contributions
must likewise be matched to the symmetrized amplitude,
which introduces an additional factor of two in their
normalization. Following the prescription of
Sec.~\ref{sec:QFTmatching}, the resulting potential is
\begin{equation}\label{eq:GGpotential}
V_{GG}(r)
=-\alpha_{GG}\frac{e^{-m_Gr}}{r}
+\beta_{GG}\delta_{r_0}^{(3)}(\boldsymbol r)
\text{ , }
\end{equation}
with reduced mass $\mu_{GG}=m_G/2$.
The attractive Yukawa interaction arises from scalar
exchange, with strength
\begin{equation}\label{eq:GGalpha}
\alpha_{GG}
=\frac{g_3^2}{16\pi m_G^2}
=\frac{25m_G^2}{16\pi\Lambda_G^2}
\text{ .}
\end{equation}
The contact interaction receives contributions from
the quartic vertex and the $s$-channel exchange.
Namely, since we focus on the onset of binding, the
$s$-channel propagator can be evaluated at the
two-body threshold  $s\simeq4m_G^2$.  and replaced by a local contact
interaction. Corrections from its energy dependence
vanish as the binding energy approaches zero, making
this approximation suitable for determining the
first-binding threshold.
Both contributions are
repulsive:
\begin{equation}\label{eq:GGbeta}
\begin{split}
\beta_4
&=\frac{g_4}{8m_G^2}
=\frac{11}{8\Lambda_G^2} \text{ , }
\beta_s=\frac{g_3^2}{24m_G^4}
=\frac{25}{24\Lambda_G^2} \text{ , }\\
\beta_{GG}
&=\beta_4+\beta_s
=\frac{29}{12\Lambda_G^2}>0 \text{ .}
\end{split}
\end{equation}
As a consistency check, the tree-level threshold
amplitude agrees with the symmetrized Born expression:
\begin{equation}\label{eq:GGcheck}
\left.\mathcal M_{GG}\right|_{s=4m_G^2,t=u=0}
=\frac{92m_G^2}{3\Lambda_G^2}
=-8m_G^2\int d^3r\,V_{GG}(r)
\text{ .}
\end{equation}

\begin{figure}[tb]
 \centering
 \includegraphics[width=0.75\linewidth]{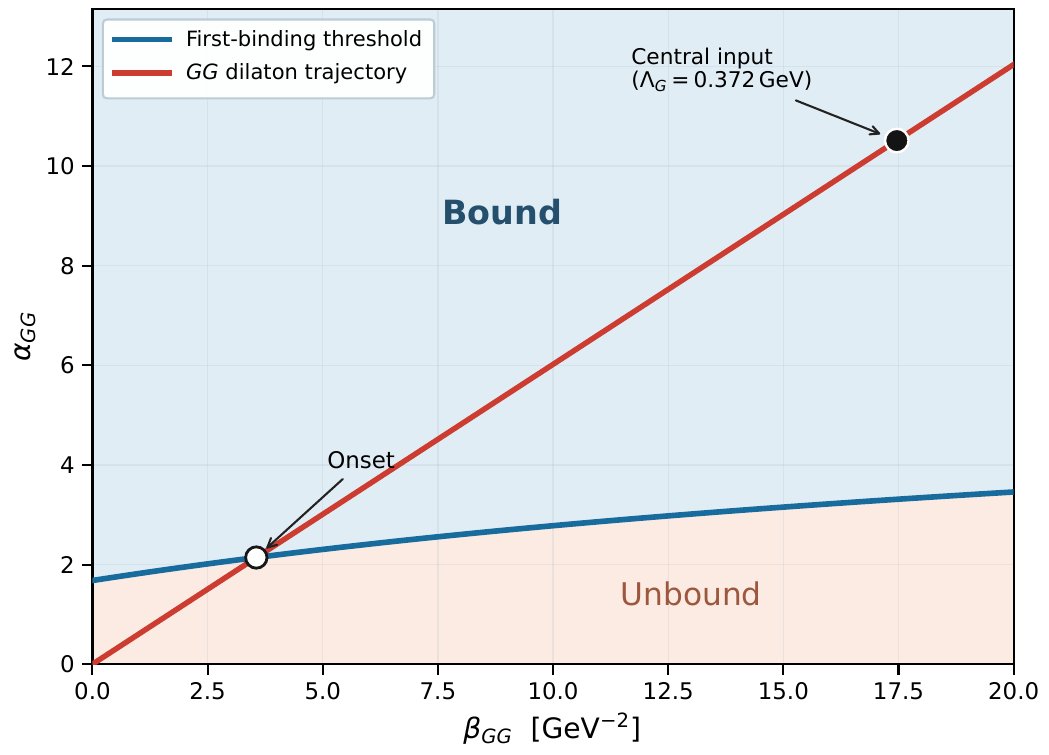}
 \caption{Scalar $GG$ first-binding condition for $m_G=1.710$ GeV and $r_0=0.30$ fm. The blue curve separates \emph{bound} (above; pale blue) and \emph{unbound} (below; pale orange) parameter sets. The red  trajectory fixed by Eq.~\eqref{eq:GGtrajectory}. The \emph{open circle} shows the first threshold, while the \emph{filled dot} shows our central lattice-matched point $\Lambda_G=0.372$ GeV. 
 %The right panel magnifies the onset; the filled dot is outside that narrower view.
 }
 \label{fig:trajectory}
\end{figure}

The Yukawa and contact strengths are not independent,
but are correlated by the dilaton relations:
\begin{equation}\label{eq:GGtrajectory}
\alpha_{GG}
=\frac{75m_G^2}{116\pi}\beta_{GG}
\text{ , }
\end{equation}
with $a_{GG}=\alpha_{GG}$ and
$b_{GG}=m_G^2\beta_{GG}$. When $m_G=1.71$ is kept fixed, the actual value of the pair $(\beta_{GG},\alpha_{GG})$ is confined to a straight line, followed by the red dilaton trajectory
in Fig.~\ref{fig:trajectory}. 
The critical value of $\Lambda_G$ is determined
by the first-binding condition introduced in
the previous section. 

For the value $r_0=0.30$ fm, corresponding
to $m_Gr_0=2.600$, the first $GG$ binding
threshold occurs at
\begin{equation}\label{eq:GGcritnumber}
\Lambda_{G,\rm crit}^{GG}
=0.824\ \mathrm{GeV} \text{ , }
\end{equation}
with $\alpha_{GG,\rm crit}=2.141$ and
$\beta_{GG,\rm crit}=3.558\ \mathrm{GeV}^{-2}$.
We recall that:
\begin{align}
   &\Lambda_{G} < \Lambda_{G,\rm crit}^{GG} \rightarrow \text{scalar glueballonium exists;} \\ &\Lambda_{G} \geq \Lambda_{G,\rm crit}^{GG} \rightarrow \text{no scalar glueballonium forms.} 
\end{align}

Without the repulsive contact interaction,
the corresponding threshold is
$\Lambda_{G,\rm crit}^{GG}=0.930$ GeV, slightly larger than the value quoted above. 
At the \emph{central} lattice-matched dilaton scale:
\begin{equation}\label{eq:GGphysicalpoint}
\begin{aligned}
 \Lambda_G&=0.372\ \mathrm{GeV},\qquad
 \alpha_{GG}=10.509,\quad\beta_{GG}=17.463\ \mathrm{GeV}^{-2},\\
 &(a_{GG},b_{GG})=(10.509,51.065).
\end{aligned}
\end{equation}
The value $\Lambda_G=0.372$ GeV is \textit{sizably} smaller (by a factor of about 2.22) than $\Lambda_{G,\rm crit}^{GG}=0.824$ GeV. Similarly, $\alpha_{GG}=10.51>\alpha_{GG,\text{crit}}=2.141$.
Visually, the black dot of Fig.~\ref{fig:trajectory}  lies deeply in the bound region. This statement concerns the \emph{existence} of at least one model bound state, but the employed nonrelativistic approach should not be used to infer its deeply bound mass at these strong couplings. Parameter variation is discussed below. 

\iffalse
The uncertainties of the dilaton parameters follow from the
lattice determinations of the scalar glueball mass and its
gluonic matrix element, as discussed above. Propagating
the uncertainties of these two lattice inputs at linear
order, and neglecting their covariance, gives
$\alpha_{GG}=10.51\pm5.44$ and
$\beta_{GG}=(17.46\pm8.05)\,\mathrm{GeV}^{-2}$.
These uncertainties are strongly correlated, since the
dilaton potential fixes
$b_{GG}=(116\pi/75)a_{GG}$.
Consequently, the attractive Yukawa and repulsive contact
strengths cannot be varied independently. The robustness
of the binding prediction should instead be assessed along
the dilaton trajectory. In particular, even the smallest
critical scale found in our scan,
$\Lambda_{G,\mathrm{crit}}\simeq0.588$ GeV, firemains
above the upper end of the nominal one-standard-deviation
interval, $\Lambda_G=0.458$ GeV.
\fi

\subsubsection{Dependence of the binding threshold on the contact radius}

An important outcome of our analysis is the robustness of the
binding prediction under variations of the contact radius $r_0$.
To illustrate this point, we repeat the calculation of the
first $GG$ binding threshold for different values of $r_0$.

\begin{figure}[tb]
\centering
\includegraphics[width=0.90\linewidth]
{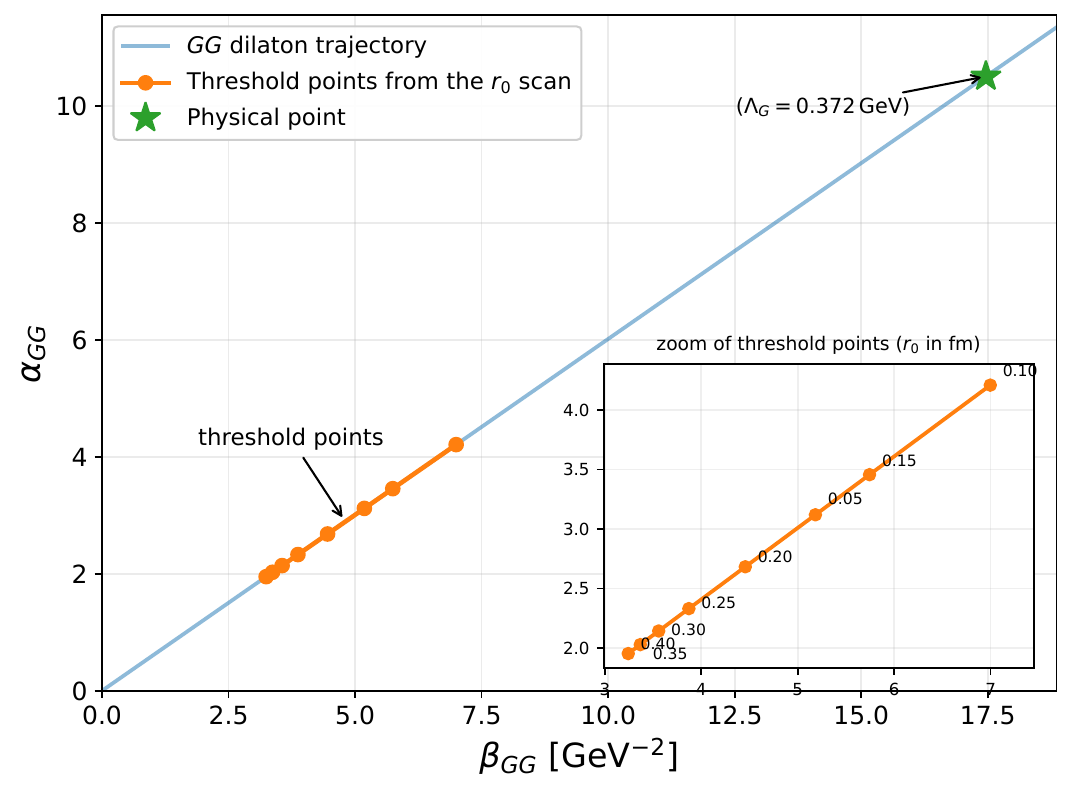}
\caption{First-binding threshold points in the
$(\beta_{GG},\alpha_{GG})$ plane for contact radii
$r_0=0.05$--$0.40$ fm at $m_G=1.710$ GeV.
The orange markers are labeled by $r_0$ in the inset.
The blue line is the dilaton trajectory, while the
star denotes the central lattice-matched point
at $\Lambda_G=0.372$ GeV.}
\label{fig:GG_radius_scan}
\end{figure}

Figure~\ref{fig:GG_radius_scan} displays the resulting
threshold points in the $(\beta_{GG},\alpha_{GG})$ plane
for eight representative radii between $0.05$ and $0.40$ fm.
All points lie on the same straight line, as required by
Eq.~\eqref{eq:GGtrajectory}, since the dilaton Lagrangian
fixes the relation between $\alpha_{GG}$ and $\beta_{GG}$ (once the glueball mass $m_G$ is kept fixed). The physical lattice-matched point is also shown
for comparison.
The inset magnifies the region containing the threshold
points, which are labeled by their corresponding values of
$r_0$.

The dependence on $r_0$ is nonmonotonic.
Among the tested $r_0$ values, the smallest critical dilaton
scale is found at $r_0=0.10$ fm, where
$\Lambda_{G,\mathrm{crit}}=0.588$ GeV,
corresponding to the largest critical Yukawa coupling,
$\alpha_{GG,\mathrm{crit}}=4.211$.
For larger radii, the critical dilaton scale increases,
reaching $\Lambda_{G,\mathrm{crit}}=0.863$ GeV
at $r_0=0.40$ fm. As discussed in Appendix~\ref{app:app},
the repulsive contact interaction becomes ineffective
in the limit $r_0\to0$, and the critical scale approaches
continuously the pure-Yukawa value
$\Lambda_{G,\mathrm{crit}}^{Y}\simeq0.930$ GeV.

The central lattice-matched parameters lie in the
bound region for all radii considered in
Fig.~\ref{fig:GG_radius_scan}. To characterize the
distance from the onset without treating the
correlated Yukawa and contact couplings as independent,
we introduce the nominal separation
$\mathcal N_\Lambda=
(\Lambda_{G,\mathrm{crit}}-\Lambda_G)/\Delta\Lambda_G$,
with $\Lambda_G=(0.372\pm0.086)$ GeV.
As summarized in Table~\ref{tab:GG_GP_nominal_separation},
$\mathcal N_\Lambda=5.26$ for the benchmark
$r_0=0.30$ fm, while it decreases to $2.51$
near $r_0=0.10$ fm, where binding is the least favored radius range among the points shown in Fig. \ref{fig:GG_radius_scan}.

\begin{table}[tb]
\centering
\caption{Critical dilaton scales for the two channels $GG$ and $GP$,
calculated at the central lattice glueball masses.
The last column gives the nominal separation
$\mathcal N_\Lambda=
(\Lambda_{G,\mathrm{crit}}-0.372\,\mathrm{GeV})/
(0.086\,\mathrm{GeV})$. Below, the least favorable entries refer to the numerical values tested in Figs. 5 and 7. The $GP$ entries refer
to the local-threshold approximation; see Sec. 3.2.}
\label{tab:GG_GP_nominal_separation}
\begin{tabular}{lccc}
\hline
Channel & $r_0$ [fm] &
$\Lambda_{G,\mathrm{crit}}$ [GeV] &
$\mathcal N_\Lambda$ \\
\hline
$GG$ (benchmark) & 0.30 & 0.8241 & 5.26 \\
$GG$ (least favorable) & 0.10 & 0.5877 & 2.51 \\
$GP$ (benchmark, local) & 0.30 & 0.7945 & 4.91 \\
$GP$ (least favorable, local) & 0.40 & 0.7924 & 4.89 \\
\hline
\end{tabular}
\end{table}

These values represent nominal separations in
units of the quoted uncertainty of $\Lambda_G$,
rather than Gaussian exclusion significances.
Indeed, $\Lambda_G$ depends nonlinearly on
the original lattice inputs, while the binding
threshold also changes with $m_G$.
Furthermore, no probability distribution for
$r_0$ is assumed, so these separations cannot
be interpreted as an overall probability
for the existence of glueballonium. Yet, the overall picture is that binding takes place for any case considered here. 

Finally, it is useful to compare this result with the earlier
relativistic studies of scalar glueballonium
\cite{Giacosa:2021brl,Trotti:2021Mass}.
Using a twice-subtracted on-shell unitarization of the
dilaton scattering amplitude, these works obtained
$\Lambda_{G,\mathrm{crit}}\simeq0.504$ GeV for
$m_G\simeq1.7$ GeV. An alternative $N/D$ treatment
in Ref.~\cite{Giacosa:2021brl} yielded the very similar
value $\Lambda_{G,\mathrm{crit}}\simeq0.500$ GeV.

Thus, these relativistic approaches also predict
scalar glueballonium for the central lattice-matched
scale $\Lambda_G=0.372$ GeV. However, their critical
values of $\Lambda_G$ are smaller than even the
minimum found in our nonrelativistic analysis.
Since binding occurs for
$\Lambda_G<\Lambda_{G,\mathrm{crit}}$, our approach,
which is particularly suited to investigating
the onset of bound-state formation, favors the
existence of scalar glueballonium over a wider
range of dilaton parameters. 
%This quantitative difference should nevertheless be interpreted in light of the different approximations and matching prescriptions employed in the two approaches \cite{Giacosa:2021brl}.

\subsubsection{Comparison with lattice scattering}
An early lattice study of pure SU(3) Yang--Mills
theory yielded an estimate of the scalar three-glueball
coupling, $g_{GGG}/m_G\simeq 50\pm 8$
\cite{deForcrand:1985}.
More recently, preliminary finite-volume results
presented at LATTICE2024
\cite{Hansen:2024Talk,Morningstar:2025Update}
were found to be compatible with this earlier estimate.
For comparison, our dilaton model predicts
$g_{GGG}/m_G=5m_G/\Lambda_G\simeq 23$,
corresponding to $\alpha_{GG}\simeq 10.5$, thus the lattice estimate is even larger.

Moreover, the HAL QCD studies by N. Yamanaka et al. extracted interglueball potentials and phase shifts, primarily in $SU(2)$ \cite{Yamanaka:2020qom,Yamanaka:2021xqh}. An exploratory $SU(3)$/$SU(4)$ calculation also exists but does not yield comparably precise scattering observables \cite{Yamanaka:2019pot}. These are valuable tests of a potential description and show substantial attraction. In the convention $V_Y^{\rm lat}(r)=Y e^{-m_\phi r}/(4\pi r)$, the dimensionless strength is $\alpha_{\rm fit}=-Y/(4\pi)$. 
Here $m_\phi$ is the scalar glueball mass in
$SU(2)$ Yang--Mills theory, determined on the lattice
to be approximately $6\Lambda$, where $\Lambda$
denotes the corresponding Yang--Mills scale. In particular, the analysis of Ref. \cite{Yamanaka:2020qom} 
gives $Y=-231(8)$, or $\alpha_{\rm fit}=18.38(64)$, while the later $SU(2)$ calculation of Chu \emph{et al.} reports $Y=-17.6(2.6)$, or $\alpha_{\rm fit}=1.40(21)$ \cite{Chu:2024Glueball}. These fitted strengths are manifestly not mutually compatible as determinations of a universal Yukawa coefficient, and they should not be averaged.

Our lattice-matched dilaton value $\alpha_{GG}\simeq10.5$ belongs qualitatively to the \emph{strong-attraction regime} indicated by the Yamanaka fit rather than the much weaker single-Yukawa Chu fit. This is encouraging \emph{qualitatively}, but there is no quantitative agreement test: both of those potentials are $SU(2)$ results, have different fitted smeared contact term forms (Chu \emph{et al.} favor a two-Gaussian fit), and depend on the chosen potential extraction or fit procedure. The true scattering observables should be scheme independent if a nonlocal HAL QCD potential and its derivative expansion are treated consistently \cite{Kawai:2017goq}. In particular, the reported cross-section magnitude constrains $|\scatt|$, not its sign, which is essential for distinguishing a shallow bound pole from a virtual one. An $SU(3)$ calculation of signed $\scatt$, effective range $r_e$, and the scalar pole residue would be the clean input for our potential matching.

A further interesting comparison can be made using
large-$N_c$ scaling. Since the three-glueball coupling
scales as $g_{GGG}\sim N_c^{-1}$, the Yukawa strength
behaves as $\alpha\sim N_c^{-2}$ at leading order
\cite{tHooft:1973alw}.
Applying this scaling to the $SU(2)$ Yukawa fit of
Ref.~\cite{Yamanaka:2020qom}, we obtain
an estimate for $SU(3)$,
\begin{equation}
\begin{aligned}
Y_{SU(3)}^{\mathrm{est}}
&\simeq \left(\frac{2}{3}\right)^2 Y_{SU(2)}
=-102.7(3.6),\\
\alpha_{SU(3)}^{\mathrm{est}}
&=-\frac{Y_{SU(3)}^{\mathrm{est}}}{4\pi}
=8.17(28) \text{ , }
\end{aligned}
\label{eq:Yamanaka_largeNc}
\end{equation}
comparable to our dilaton prediction
$\alpha_{GG}\simeq10.51$.

In summary, the results and comparisons above are promising, but novel and more precise  $GG$ scattering parameters on the lattice are needed. 
%At the same time, the precise determination of all $J^{
%In summary, all estimates above point towards a sizable scalar glueball self-interaction. Yet, a direct lattice determination of the $GG$ scattering amplitude is still needed to establish whether a glueballonium bound state exists.

\subsection{Scalar--pseudoscalar glueballonium:
\texorpdfstring{$GP$}{GP}}\label{sec:GP}

In addition to the scalar dilaton/glueball field $G$,
we introduce a pseudoscalar glueball $P$.
A minimal potential, including an independent
stabilizing pseudoscalar self-interaction, is
\cite{Giacosa:2026theta}:
\begin{equation}\label{eq:fullpotential}
V(G,P)=V_{\rm dil}(G)
+\frac{\lambda_{GP}}{2}G^2P^2
+\frac{\lambda_P}{4!}P^4
\text{ , }
\lambda_{GP}>0 \text{ , } \lambda_P\geq0
\text{ .}
\end{equation}
Lattice spectroscopy places a pseudoscalar $0^{-+}$
glueball among the low-lying pure-gauge states 
\cite{Chen:2005mg}, motivating a second molecular channel.

The pseudoscalar glueball and its couplings to the
axial anomaly and ordinary mesons have a substantial
effective-theory literature 
\cite{Rosenzweig:1981cu,Masoni:2006rz,  Eshraim:2012jv,Giacosa:2025dvu}.
As a recent experimental development, it is worth
mentioning the pseudoscalar resonance $X(2370)$,
observed by BESIII and exhibiting striking
glueball-like features
\cite{BESIII:2010gmv,BESIII:2024SpinParity,BESIII:2026NewModes,BESIII:2026Dominant}.
Remarkably, its decays can be understood within
a theory implementing fractional 't Hooft
determinants \cite{Giacosa:2026theta}.

Shifting $G\mapsto\Lambda_G+G$, Eq. (\ref{eq:fullpotential}) generates
\begin{equation}\label{eq:GPvertices}
\begin{split}
m_P^2&=\lambda_{GP}\Lambda_G^2
\text{ , }
h=2\lambda_{GP}\Lambda_G
=\frac{2m_P^2}{\Lambda_G}
\text{ , }\\
\mathcal L_{\rm int}
&\supset-\frac h2 GP^2
-\frac{\lambda_{GP}}2G^2P^2
-\frac{\lambda_P}{4!}P^4
\text{ .}
\end{split}
\end{equation}
The mass relation assumes no independent bare
$P$ mass term, in agreement with dilatation symmetry.
The $P^4$ interaction does not
contribute directly to $GP\to GP$ at tree level,
but enters $PP\to PP$ and ensures that the
potential is bounded from below along the
$P$ direction for $\lambda_P\geq0$.
Numerically, we use $m_P=2.560$ GeV \cite{Chen:2005mg}.

\begin{figure}[tb]
\centering
\includegraphics[width=\linewidth]{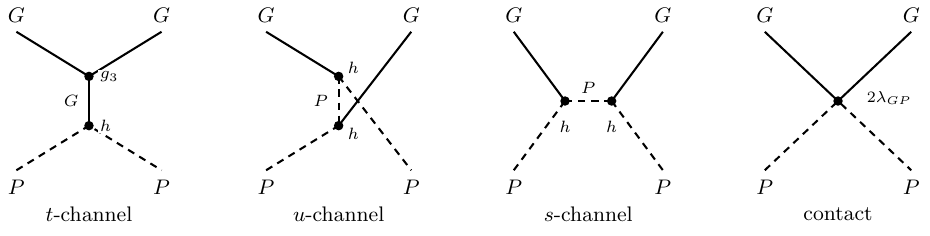}
\caption{The $GP\to GP$ tree diagrams:
$t$-channel scalar exchange, $u$- and
$s$-channel pseudoscalar exchange,
and the local $G^2P^2$ vertex.}
\label{fig:GP_diagrams}
\end{figure}

The $GP$ tree diagrams are shown in
Fig.~\ref{fig:GP_diagrams}.
Since the two external particles are
distinguishable, the $u$-channel exchange is
a genuine additional contribution, rather than
the Bose symmetrization of the $t$ channel.
The invariant amplitude reads
\begin{equation}\label{eq:GPamp}
\mc_{GP}
=-\frac{g_3h}{t-m_G^2}
-\frac{h^2}{u-m_P^2}
-\frac{h^2}{s-m_P^2}
-2\lambda_{GP}
\text{ .}
\end{equation}

Matching the $t$-channel contribution to the
nonrelativistic potential, without
identical-particle symmetrization, gives
the Yukawa coefficient and reduced mass
\begin{equation}\label{eq:GPalpha}
\alpha_{GP}
=\frac{g_3h}{16\pi m_Gm_P}
=\frac{5m_Gm_P}{8\pi\Lambda_G^2}
\text{ , }
\mu_{GP}=\frac{m_Gm_P}{m_G+m_P}
\text{ .}
\end{equation}

The remaining three diagrams are approximated
by a local interaction evaluated at the elastic
threshold, where $s=(m_G+m_P)^2$ and
$u=(m_G-m_P)^2$. We obtain
\begin{align}\label{eq:GPbeta}
\beta_{GP}
&=\beta_u+\beta_s+\beta_4 \nonumber\\
&=\frac{1}{4m_Gm_P}\left[
-\frac{h^2}{m_G(2m_P-m_G)}
+\frac{h^2}{m_G(m_G+2m_P)}
+2\lambda_{GP}
\right]\nonumber\\
&=-\frac{m_Gm_P}
{2(4m_P^2-m_G^2)\Lambda_G^2}<0
\text{ .}
\end{align}
Here $\beta_u<0$ is the attractive crossed
$u$-channel contribution, whereas
$\beta_s>0$ and $\beta_4>0$ arise from
the repulsive $s$-channel exchange and
the four-point vertex, respectively.
For the masses considered here,
the crossed exchange slightly overcompensates
the two repulsive terms, resulting in
a net attractive contact interaction.

The resulting potential in the local-threshold
approximation is therefore
\begin{equation}\label{eq:GPpot}
V_{GP}(r)
=-\alpha_{GP}\frac{e^{-m_Gr}}{r}
+\beta_{GP}\delta_{r_0}^{(3)}(\boldsymbol r)
\text{ .}
\end{equation}

The local-threshold approximation does
not necessarily reproduce the effects of retaining
the full momentum dependence of crossed exchange.
A related distinction arises in other contexts
\cite{Xu:2022Neutrino}, although that analysis does
not provide an additional matching condition for
the present system.
The alternative treatment retaining the finite-range
crossed pseudoscalar exchange is examined in
Appendix~\ref{app:app}, where we find that
the central binding prediction survives, although
the quantitative thresholds change slightly.

At $r_0=0.30$ fm, the first $GP$ binding threshold
occurs at
\begin{equation}\label{eq:GPcrit}
\Lambda_{G,\rm crit}^{GP}=0.795\ \mathrm{GeV}
\text{ , }
\end{equation}
corresponding to $\alpha_{GP,\rm crit}=1.3795$
and $\beta_{GP,\rm crit}=-0.1489\ \mathrm{GeV}^{-2}$.
At the central lattice-matched scale
$\Lambda_G=0.372<\Lambda_{G,\rm crit}$ GeV, well within the bound region of the model. 
The corresponding values read: 
\begin{equation}\label{eq:GPcentral}
\alpha_{GP}=6.293 \text{ , }
\beta_{GP}=-0.6791\ \mathrm{GeV}^{-2}
\text{ , }
(a_{GP},b_{GP})=(7.546,-2.381)
\text{ .}
\end{equation}

\begin{figure}[tb]
\centering
\includegraphics[width=0.90\linewidth]
{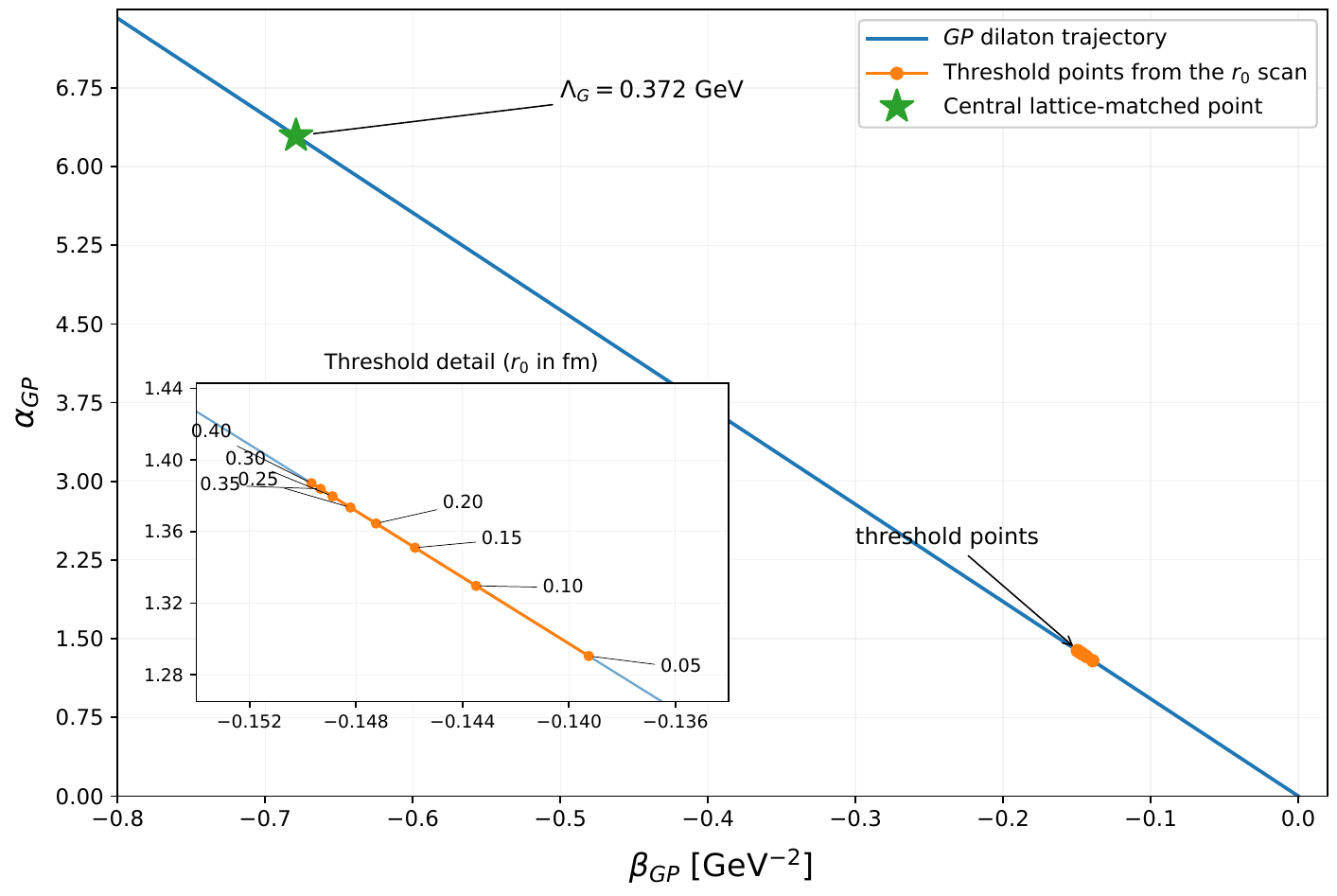}
\caption{First-binding threshold points for the $GP$
potential in the $(\beta_{GP},\alpha_{GP})$ plane,
for $r_0=0.05$--$0.40$ fm and the central lattice
masses $m_G=1.710$ GeV and $m_P=2.560$ GeV.
The calculation uses the local-threshold approximation
for the short-range interaction.
The orange markers, labeled by $r_0$ in the inset,
lie along the dilaton trajectory. The star denotes
the central lattice-matched point at
$\Lambda_G=0.372$ GeV. Notice the very weak dependence on $r_0$.}
\label{fig:GP_radius_scan}
\end{figure}

The $GP$ channel behaves differently from $GG$:
the first-binding threshold
depends only \textit{weakly} on the contact radius.
As shown in Fig.~\ref{fig:GP_radius_scan},
$\Lambda_{G,\rm crit}^{GP}$ decreases from
$0.8215$ to $0.7924$ GeV as $r_0$ increases
from $0.05$ to $0.40$ fm.
The corresponding nominal separations from
$\Lambda_G=(0.372\pm0.086)$ GeV are approximately
$4.9$ quoted standard uncertainties
(Table~\ref{tab:GG_GP_nominal_separation}),
with the same statistical qualifications
discussed in the $GG$ channel.

In fact, for $\beta_{GP}<0$, it is sufficient
to verify whether the Yukawa interaction alone
supports a bound state. Its dimensionless
strength at the central parameters is
$a_{GP}=7.546$, considerably larger than the
pure-Yukawa critical value $a_Y\simeq1.6798$.
Hence, a $GP$ bound state exists even in the
absence of the contact interaction. Adding
the attractive contact term can only reinforce
binding, independently of its radius or
spatial profile (see App. \ref{app:app} where the treatment beyond the local approximation is discussed).

\iffalse
This conclusion also remains valid under an
illustrative propagation of the lattice-input
uncertainties. In addition to the scalar inputs
of Eq.~\eqref{eq:GG_lattice_inputs}, we use
$m_P=(2.560\pm0.125)$ GeV, combining the two
uncertainties quoted by Chen \emph{et al.}
\cite{Chen:2005mg} in quadrature.
Treating these three inputs as independent Gaussian
variables, and recalculating $\Lambda_G$ consistently
from the scalar mass and matrix element, we find
$a_{GP}\gtrsim2.23$ throughout the nominal
$95.45\%$ joint input region
($\Delta\chi^2=8.02$ for three parameters).
This remains above the pure-Yukawa threshold
$a_Y\simeq1.6798$. Thus, within the present
potential model, $GP$ binding does not rely on
the detailed short-range interaction.
The quoted uncertainty region neglects correlations
between the lattice inputs and does not include
systematic uncertainties from momentum-dependent
interactions, form factors, or coupled channels.
\fi

An $S$-wave $GP$ molecule has $J^{PC}=0^{-+}$
and mass $m_G+m_P-B$. In parity-conserving
pure Yang--Mills theory, it cannot decay into
two $0^{++}$ glueballs: $J=0$ requires relative
$L=0$, implying positive parity. The $GT$
final state also has positive parity for $J=0$,
since $L=2$ is required. Furthermore, for the
Chen \emph{et al.} central masses, the $TT$
and $GGG$ thresholds, approximately $4.78$
and $5.13$ GeV, lie above the $GP$ threshold
of $4.27$ GeV. Thus, a $GP$ molecule
is stable against strong decay in pure
Yang--Mills theory without CP-breaking terms.

\subsection{What about other glueballonia?
\texorpdfstring{$PP$, $GT$, and $TT$}{PP, GT, and TT}}
\label{sec:other}

The strong scalar-exchange coupling motivates the
investigation of further glueball pairs. We treat the
constituent glueballs as stable pure-gauge degrees of
freedom and study binding relative to their two-body
thresholds. Such bound configurations may nevertheless
couple to lighter channels and appear as resonances.

\paragraph{$PP\to PP$.}
For two identical pseudoscalars, the $hGP^2$
vertex with $h
=2m_P^2/\Lambda_G$ generates scalar exchange in the $t$,
$u$, and $s$ channels, while the quartic
interaction $\lambda_P P^4/4!$ provides an
additional contact contribution.
The tree-level amplitude is
\begin{equation}\label{eq:PPamp}
\mc_{PP}
=-\frac{h^2}{t-m_G^2}
-\frac{h^2}{u-m_G^2}
-\frac{h^2}{s-m_G^2}
-\lambda_P
\text{ .}
\end{equation}
As for $GG$, the $t$- and $u$-channel contributions
correspond to a single Bose-symmetrized Yukawa
potential. The resulting coefficients are
\begin{equation}\label{eq:PPcoeff}
\begin{split}
\alpha_{PP}
=\frac{h^2}{16\pi m_P^2}
=\frac{m_P^2}{4\pi\Lambda_G^2}
\text{ , }
\beta_{PP}
=\frac{\lambda_P}{8m_P^2}
+\frac{h^2}{8m_P^2(4m_P^2-m_G^2)}
\text{ , }
\mu_{PP}&=\frac{m_P}{2}
\text{ .}
\end{split}
\end{equation}
Both contributions to $\beta_{PP}$ are repulsive
for $\lambda_P\geq0$. At $\Lambda_G=0.372$ GeV,
we obtain $\alpha_{PP}=3.769$ and
$a_{PP}=m_P\alpha_{PP}/m_G=5.642$.
For $\lambda_P=0$, the Yukawa attraction is
sufficiently strong to generate
a bound state.
For $r_0=0.30$ fm, our model predicts binding
up to $\lambda_{P,\rm crit}\simeq3.1\times10^3$.
This large formal value should not
be interpreted as a quantitatively reliable
bound on the physical self-coupling, but merely shows that the contact repulsion should be large to prevent binding\footnote{A related preliminary
relativistic investigation of the $PP$ channel,
following the approach of
Ref.~\cite{Giacosa:2021brl}, was carried out
in Ref.~\cite{Pniak:2026MSc} using the same
effective interactions. This approach shows that binding takes place for $\lambda_P \le 300$.}. 
An $S$-wave $PP$ state has $J^{PC}=0^{++}$
and can decay into two scalar glueballs.
It would therefore generally appear as a
resonance rather than a stable glueballonium.

\paragraph{$GT\to GT$.}
We now introduce the tensor glueball
$T_{\mu\nu}$ with $J^{PC}=2^{++}$.
Unlike a scalar field, a massive spin-2 field
has five physical polarization states and
requires an appropriate tensor kinetic
structure. As a minimal mass-generating
interaction, we consider
\begin{equation}\label{eq:GTTansatz}
\mathcal L_{GT}\supset
-\frac{\lambda_{GT}}{2}G^2
\left[
T_{\mu\nu}T^{\mu\nu}
-(T^\mu{}_\mu)^2
\right]
\text{ , }
\lambda_{GT}>0
\text{ .}
\end{equation}
The quadratic tensor structure has the
Fierz--Pauli form required for a free massive
spin-two field \cite{Koenigstein:2015asa}.
The trace term vanishes for physical on-shell
tensor polarizations.
Above, $G$ denotes the unshifted dilaton field. For physical tensor
polarizations the trace vanishes.
After shifting $G\mapsto\Lambda_G+G$, we find
\begin{equation}
m_T^2=\lambda_{GT}\Lambda_G^2
\text{ , }
h_T=2\lambda_{GT}\Lambda_G
=\frac{2m_T^2}{\Lambda_G}
\text{ ,}
\end{equation}
where $h_T$ is defined through the trilinear
interaction
$-h_T G T_{\mu\nu}T^{\mu\nu}/2$
in the physical tensor sector.

The leading $t$-channel scalar exchange is
spin diagonal in the nonrelativistic limit.
It generates an attractive Yukawa interaction
with strength
\begin{equation}\label{eq:GTalpha}
\alpha_{GT}^{(G)}
=\frac{g_3h_T}{16\pi m_Gm_T}
=\frac{5m_Gm_T}{8\pi\Lambda_G^2}
\text{ , }
\mu_{GT}=\frac{m_Gm_T}{m_G+m_T}
\text{ .}
\end{equation}
With $m_T=2.390$ GeV and
$\Lambda_G=0.372$ GeV, this gives
$\alpha_{GT}^{(G)}=5.875$ and
$a_{GT}^{(G)}=6.850$, considerably above
the pure-Yukawa threshold $a_Y\simeq1.680$.

This illustrates the strong attraction generated
by scalar exchange, but does not establish
a complete $GT$ bound-state pole.
Indeed, the same interaction in
Eq.~\eqref{eq:GTTansatz} also generates a
local $G^2T^2$ vertex, while the trilinear
$GTT$ vertex contributes through crossed
tensor exchange. Their proper treatment
requires the tensor propagator and its
spin projectors. Additional derivative
interactions and coupled channels may also
modify the result.

An $S$-wave $GT$ configuration has
$J^{PC}=2^{++}$. Since
$m_G+m_T=4.10$ GeV exceeds
$2m_G=3.42$ GeV, decay into two scalar
glueballs in a $D$ wave is kinematically
allowed\footnote{The decay $GT\to GG$ is allowed by quantum numbers
and kinematics, but requires interactions odd in
$T_{\mu\nu}$, which are absent from the minimal
Ansatz in Eq.~\eqref{eq:GTTansatz}.}. Thus, a corresponding state
would generally be resonant. 

\paragraph{$TT\to TT$.}
The interaction introduced in Eq. (\ref{eq:GTTansatz}) also
determines the coupling of two tensor
glueballs to the scalar glueball. Hence,
it naturally leads to $TT$ scattering
through scalar exchange in the $t$,
$u$, and $s$ channels.

The tensor nature of the particles is now
essential. For two identical spin-2 particles
in an $S$ wave, Bose symmetry allows total
spin $J=0,2,4$. Following the spin
decomposition discussed in
Ref.~\cite{Trotti:2022knd}, the leading
nonrelativistic amplitudes take the form
\begin{equation}\label{eq:TTamp}
\mc_{TT}^{(J)}
=-\frac{h_T^2}{t-m_G^2}
-\frac{h_T^2}{u-m_G^2}
-\frac{5\delta_{J0}h_T^2}{s-m_G^2}
\text{ , }
J=0,2,4
\text{ .}
\end{equation}
Thus, the $s$-channel contribution appears
only for $J=0$, where it is repulsive near
threshold. For $J=2,4$, the leading
interaction consists entirely of the
$t$- and $u$-channel scalar exchanges,
represented by a single Bose-symmetrized
Yukawa potential.

The corresponding Yukawa strength is
\begin{equation}\label{eq:TTalpha}
\alpha_{TT}
=\frac{h_T^2}{16\pi m_T^2}
=\frac{m_T^2}{4\pi\Lambda_G^2}
\text{ , }
\mu_{TT}=\frac{m_T}{2}
\text{ .}
\end{equation}
For the central parameters, we obtain
$\alpha_{TT}=3.285$ and
$a_{TT}=m_T\alpha_{TT}/m_G=4.591$.
Since $a_{TT}>a_Y\simeq1.680$, the
minimal scalar-exchange potential supports
bound states in the $J=2,4$ channels.

The $J=0$ channel additionally receives
a repulsive local contribution from
$s$-channel scalar exchange,
\begin{equation}\label{eq:TTbeta}
\beta_{TT}^{(0)}
=\frac{5h_T^2}
{8m_T^2(4m_T^2-m_G^2)}>0
\text{ .}
\end{equation}
Its effect on binding depends on the
short-range realization of this interaction.

These observations concern the minimal
scalar-exchange model. An independent
$T^4$ interaction, as well as further
derivative and tensor couplings, is not
fixed by Eq.~\eqref{eq:GTTansatz} and may
modify the binding conditions.
Moreover, $S$-wave $TT$ states have
$J^{PC}=0^{++},2^{++},4^{++}$ and lie
above the two-scalar threshold.
They may therefore decay into $GG$
through the corresponding even partial
waves when suitable couplings are present.
A complete description would require
the associated coupled-channel analysis.

\subsection{A comparison: the absence of Higgsonium
in the Standard Model}
\label{sec:Higgsonium}

As a final application, it is interesting to consider
the interaction between two Higgs bosons within the
Standard Model (SM) \cite{djuadi2005gi}. The possible formation of a
Higgs--Higgs bound state, called Higgsonium, was recently studied
in Ref.~\cite{Shastry:2022goq} using relativistic
unitarization methods, finding no bound state for the
SM self-couplings. For earlier studies, see Refs. \cite{Cahn:1983vi,Grifols:1991gw,DiLeo:1994xc}.

The Higgs potential, expanded around its minimum,
takes the form
\begin{equation}
V_{\mathrm{SM}}(H)
=\frac{m_H^2}{2}H^2
+\frac{g_H}{3!}H^3
+\frac{\lambda_H}{4!}H^4,
\qquad
g_H=\frac{3m_H^2}{v},
\quad
\lambda_H=\frac{3m_H^2}{v^2},
\label{eq:Higgs_potential}
\end{equation}
where $H$ denotes the physical Higgs fluctuation,
$m_H\simeq125$ GeV, and $v\simeq246$ GeV.

Applying the same nonrelativistic matching as for the
$GG$ system, we obtain
\begin{equation}
V_{HH}(r)=
-\alpha_{HH}\frac{e^{-m_Hr}}{r}
+\beta_{HH}\delta_{r_0}^{(3)}(\mathbf r),
\label{eq:Higgs_pair_potential}
\end{equation}
with
\begin{equation}
\alpha_{HH}
=\frac{g_H^2}{16\pi m_H^2}
=\frac{9m_H^2}{16\pi v^2}
\simeq0.0462,
\qquad
\beta_{HH}
=\frac{\lambda_H}{8m_H^2}
+\frac{g_H^2}{24m_H^4}
=\frac{3}{4v^2}>0.
\label{eq:Higgs_alpha_beta}
\end{equation}
As for the $GG$ case, the $s$-channel exchange and the quartic vertex
both contribute to the repulsive contact interaction,
while the $t$- and $u$-channel exchange generate the
Bose-symmetrized Yukawa attraction.

Since the exchanged particle and the constituents
have the same mass, $\mu_{HH}=m_H/2$, and the
dimensionless Yukawa strength is simply
$a_{HH}=\alpha_{HH}$. This value is much smaller
than the pure-Yukawa critical coupling
$a_Y\simeq1.6798$. Hence, even in the absence of
the repulsive contact term, the attraction is far
too weak to produce a bound state. Including the
positive contact interaction can only make binding
more difficult, independently of its nonnegative
normalized spatial profile.

Thus, within the present potential framework,
the Standard Model does not predict Higgsonium,
in agreement with the relativistic analysis of
Ref.~\cite{Shastry:2022goq}. In contrast,
the dilaton self-interactions generate a much
stronger Yukawa attraction between scalar glueballs.
Extensions of the SM with enhanced Higgs
self-couplings may nevertheless allow Higgsonium,
as investigated in Ref.~\cite{Shastry:2022goq}.

\section{Conclusions}\label{sec:conclusions}

In this work, we have investigated the possible formation
of glueballonia using a nonrelativistic potential approach,
with particular attention to the onset of bound-state
formation. The attractive Yukawa interaction generated
by scalar-glueball exchange is supplemented by a smeared contact term. We have established the conditions for the onset
of binding and examined the role of the finite range of the
contact interaction, including the subtleties of its
zero-range limit in three spatial dimensions.

For two scalar glueballs, the parameters of the NR potential are fixed through matching with the dilaton effective Lagrangian. The main result is clear:
at the central lattice-matched parameters, the dilaton
potential supports at least one bound state, the scalar
glueballonium. This conclusion is robust under the
investigated variations of the range of the repulsive
interaction. The uncertainty analysis
also identifies where this conclusion becomes sensitive
to the lattice inputs and the short-range interaction.

The same approach, applied to an extension of the dilaton potential that incorporates the pseudoscalar glueball, favors the formation of a pseudoscalar
glueballonium made of a scalar and a pseudoscalar glueball.
In the local-threshold approximation, the net contact
interaction is attractive. The Yukawa interaction alone
already supports a bound state, so adding the contact
term can only reinforce binding, independently of its
radius or nonnegative spatial profile. This conclusion
also survives the input variations considered in our
uncertainty analysis. 

Interestingly, both $GG$ and $GP$ glueballonia states, if existent, are stable in CP-conserving YM vacuum. 

Strong scalar-exchange attraction is also found in the
$PP$, $GT$, and $TT$ systems. These channels offer further
possibilities for glueballonium formation, although their
near-threshold configurations have open decay channels
and would generally appear as resonances, even in pure gauge. Establishing
their properties requires a more complete treatment
of the short-range interactions, tensor spin structure,
and coupled channels. 

As a comparison and cross-check, we have also
considered the Standard Model Higgs boson. In that case,
the Yukawa attraction is far too weak to form a
Higgs--Higgs bound state, in agreement with previous outcomes on the subject.

Our results therefore motivate further investigations
of scalar and pseudoscalar glueballonia in pure
Yang--Mills theory, particularly through lattice
studies along two complementary directions.

(i) Determinations of low-energy glueball scattering
amplitudes in $SU(3)$ would provide a direct test
of the predicted attraction and the possible
formation of bound states. In parallel, a better
understanding of finite-volume effects, the momentum
dependence of the interactions, and coupled channels
would allow more quantitative predictions for the
properties of these states.

(ii) A detailed investigation of the excited spectrum
in the scalar and pseudoscalar channels could provide
a complementary way to identify glueballonia.
Indeed, if they exist, these states belong to the
spectrum of pure Yang--Mills theory and should appear
in lattice calculations, just as conventional
glueballs do. In particular, optimized two-glueball
interpolating operators could be constructed to
enhance their overlap with molecular configurations,
facilitating the identification of possible
additional states and the study of their
finite-volume behavior.

More generally, confirming glueballonia on the
lattice would provide a new test of effective
descriptions of gluodynamics and help constrain
their underlying parameters more precisely.

We conclude with a more general question: is a glueballonium
itself a glueball? Actually, yes, since it is also a bound
state made entirely of gluons. Yet, its nature is different,
since it is a molecular bound state of two (or more) glueballs.
It is thus a nonconventional glueball state.
This distinction becomes particularly clear when varying
the number of colors $N_c$. While conventional glueballs
survive at large $N_c$, molecular glueballonia disappear
abruptly from the bound-state spectrum once the attraction
falls below its critical strength, and do not survive in
the large-$N_c$ limit \cite{Giacosa:2024scx}.
As the critical value of $N_c$ is approached from the
binding side, the glueballonium mass tends toward the
corresponding two-glueball threshold.
Following this evolution on the lattice would offer
an additional way to distinguish molecular states
from conventional glueballs. Thus, glueballonia
would be a genuine, but intrinsically finite-$N_c$,
feature of the Yang--Mills spectrum.

%Indeed, both $\alpha_{GG}$ and $\beta_{GG}$ scale as$N_c^{-2}$. Extrapolating our central $N_c=3$ resultsgives $\alpha_{GG}(N_c)\simeq10.51(3/N_c)^2$, with the same scaling for $\beta_{GG}$.Assuming approximately $N_c$-independent glueballmasses and interaction range, binding disappearsbetween $N_c=6$ and $N_c=7$ for the benchmark$r_0=0.30$ fm. Moreover, even without the repulsivecontact interaction, the pure-Yukawa threshold$\alpha_{GG,\mathrm{crit}}\simeq1.68$ precludesbinding for $N_c\geq8$ within this leading scalingapproximation. 

\bigskip

\textbf{Acknowledgments} We thank E. Trotti, N. Yamanaka, R. Porowski for a preliminary discussion and study on this subject. We also thank L. Tinti and M. Rohrmoser for useful discussions, and A. Pniak for her comments on the $PP$ case. 
This work
was supported by the Polish Minister of Science under the ‘Regional Excellence Initiative’
program (project RID/SP/0015/2024/01).
 ChatGPT (OpenAI) was used to assist in refining the
authors ideas, revising the language and providing feedback. The scientific ideas and results are those of
the author.

\appendix

\section{Smeared contact term and GP beyond local approximation}\label{app:app}

\subsection{Dependence on \texorpdfstring{$r_0$}{r0}}

\begin{figure}[tb]
 \centering\includegraphics[width=.80\linewidth]{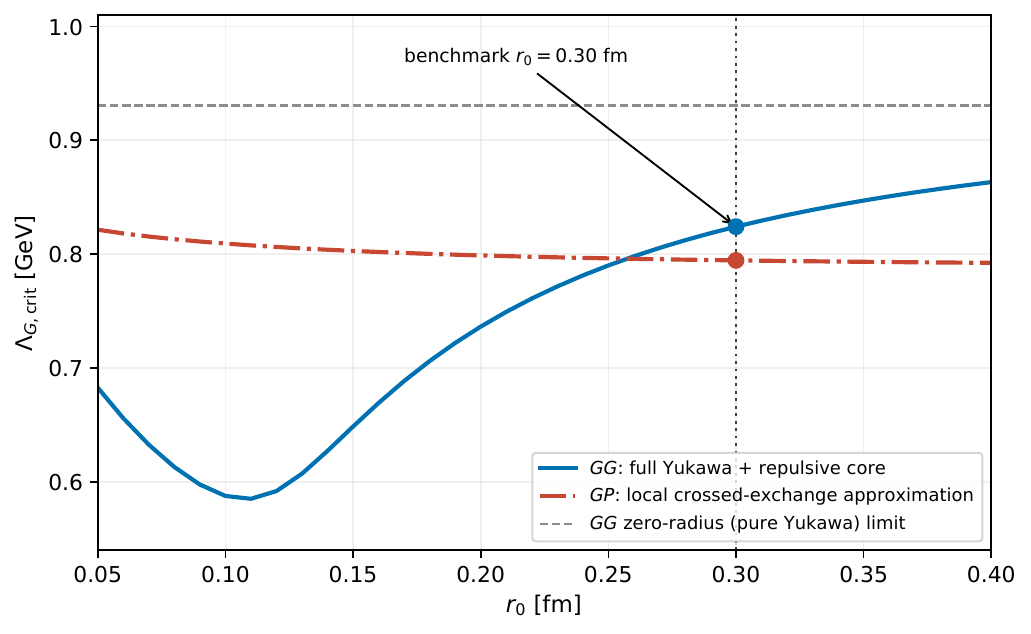}
 \caption{Sensitivity of the critical dilaton scale to the finite-core radius for the $GG$ and local $GP$ models. At $r_0\approx0.10$ fm, the $GG$ first-threshold scale has a pronounced minimum. The benchmark $r_0=0.30$ fm is marked by the dotted vertical line. The horizontal reference is the $GG$ limit as $r_0\to0$, namely the pure Yukawa critical $\Lambda_{G,\rm crit}^{Y}\simeq0.930$ GeV.}
 \label{fig:radscan}
\end{figure}

The finite-radius calculation treats $r_0$ as a physical range of the regulated potential rather than a removable ultraviolet cutoff. To make its influence transparent we hold the tree-level dilaton relations fixed and repeat the \emph{first-threshold} calculation for $r_0=0.05$--$0.40$ fm in steps of $0.01$ fm. Selected results are given in Table~\ref{tab:scan}, with the continuous scan in Fig.~\ref{fig:radscan}.
The nonmonotonicity in the $GG$ column is not a numerical artifact: as the repulsive core shrinks from a large radius, its central height initially grows and can make the Yukawa interaction less effective, but at sufficiently small $r_0$ its physical reach disappears. 
Within the local approximation, the $GP$ contact interaction is attractive,
$\beta_{GP}<0$. At fixed negative $\beta_{GP}$,
its shrinking spatial range leads to an increasingly
deep potential well, which eventually supports
bound states even without Yukawa attraction.
Consequently, unlike the repulsive case,
the critical dilaton scale does not approach
the pure-Yukawa limit as $r_0\to0$.
Instead, within the local model,
$\Lambda_{G,\mathrm{crit}}^{GP}\propto r_0^{-1/2}$
asymptotically. This formal limit should not
be regarded as a physical prediction, since
the local approximation to crossed exchange
is not controlled at arbitrarily small radii.

Note, the $GP$ result beyond local approximation is also nonmonotonic and is discussed in A.3.

 %For this reason neither curve should be extrapolated outside its stated finite-range matching assumptions without renormalization.

\subsection{Different representations of the smeared contact term} 
Next, we turn to the representation of the contact term. In the main text we regularized the short-range
interaction by means of the normalized top-hat
distribution.
One may, in addition, consider other normalized
realizations of a short-range interaction, such as
a Gaussian or an exponential form. Summarizing:
\begin{equation}
\delta^{(3)}_{r_0,T}(r)
=
\frac{3}{4\pi r_0^3}\Theta(r_0-r) \text{ , }
\delta^{(3)}_{r_0,G}(r)
=
\frac{1}{\pi^{3/2}r_0^3}
e^{-r^2/r_0^2} \text{ , }
\delta^{(3)}_{r_0,E}(r)
=
\frac{1}{8\pi r_0^3}
e^{-r/r_0}.
\end{equation}
For a fixed value of $r_0$, the corresponding
potentials read
\begin{equation}
V_i(r)=
-\alpha\frac{e^{-mr}}{r}
+\beta\delta^{(3)}_{r_0,i}(r) \text{ with } \int d^3r\,\delta^{(3)}_{r_0,i}(r)=1 \text{ for }
i=T,G,E \text{ .}
\end{equation}
One may determine, as in the main text,
the critical Yukawa strength $\alpha_{\rm crit}$
for the onset of the first $S$-wave bound state.

\begin{table}[tb]
\centering
\small
\begin{tabular}{rrrrr}
\toprule
$r_0$ [fm] & $x_0=m_Gr_0$ &
\multicolumn{3}{c}{$\Lambda_{G,\rm crit}$ [GeV]} \\
\cmidrule(lr){3-5}
& & $GG$ & $GP$ (local) & $GP$ (crossed) \\
\midrule
0.05 & 0.433 & 0.6828 & 0.8215 & 0.7384 \\
0.10 & 0.867 & 0.5877 & 0.8093 & 0.7093 \\
0.15 & 1.300 & 0.6486 & 0.8029 & 0.8075 \\
0.20 & 1.733 & 0.7364 & 0.7989 & 0.8716 \\
0.25 & 2.166 & 0.7901 & 0.7963 & 0.9092 \\
0.30 & 2.600 & 0.8241 & 0.7945 & 0.9329 \\
0.35 & 3.033 & 0.8471 & 0.7933 & 0.9488 \\
0.40 & 3.466 & 0.8633 & 0.7924 & 0.9600 \\
\bottomrule
\end{tabular}
\caption{First-binding thresholds for the $GG$ and
$GP$ systems at different contact radii, using the
central lattice glueball masses of Chen \emph{et al.}.
For $GP$, the local approximation combines the
short-distance contributions into a single contact
term, whereas the crossed treatment retains the
finite-range $u$-channel pseudoscalar exchange.
Both calculations use the same normalized top-hat
profile for their respective contact interactions.}
\label{tab:scan}
\end{table}

\begin{figure}[tb]
\centering
\includegraphics[width=0.78\textwidth]{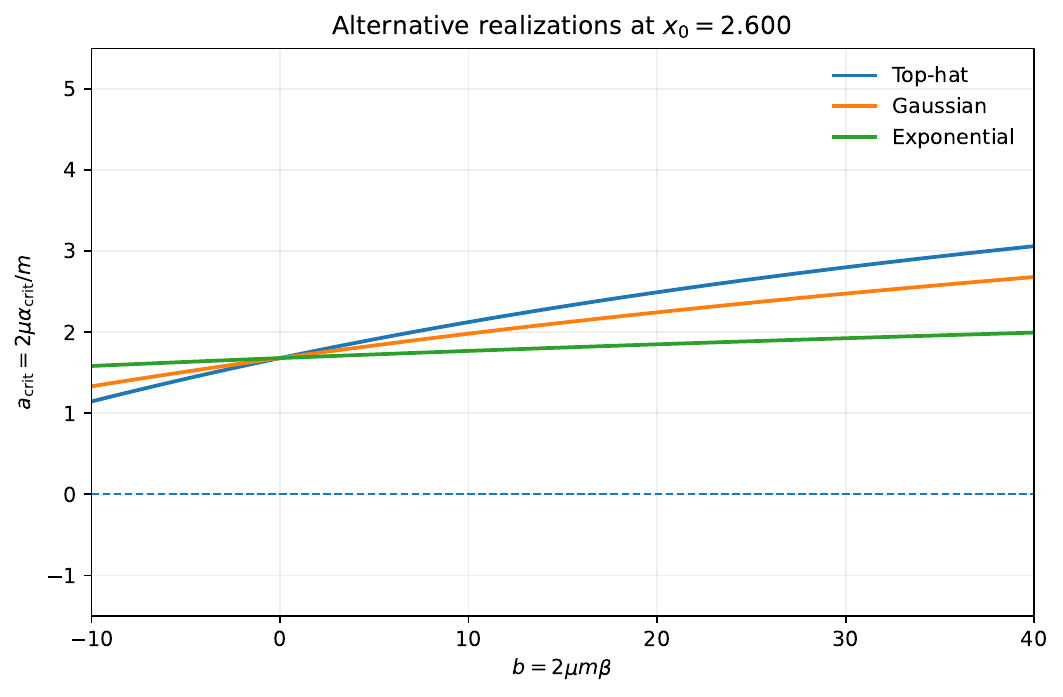}
\caption{Critical Yukawa coupling $a_{\mathrm{crit}}=2\mu\alpha_{\mathrm{crit}}/m$
as a function of $b=2\mu m\beta$ for three normalized realizations of the smeared contact
interaction at fixed $x_0=mr_0=2.600$: top-hat, Gaussian, and exponential. The three curves
coincide at $b=0$, where the contact term is absent.}
\label{fig:delta_realizations}
\end{figure}

In Fig.~\ref{fig:delta_realizations} we compare
the resulting critical curves for the three
realizations at the illustrative value
$x_0=mr_0=2.6$. As expected, all curves meet at $b=0$, where the
contact term vanishes and one recovers the
pure-Yukawa threshold. For a repulsive contact
interaction ($b>0$), the critical Yukawa strength
shows a mild dependence on the shape of the regularized contact.
At the common nominal width $r_0$ used here,
the top-hat profile produces the largest
critical attraction, followed by the Gaussian
and the exponential profiles. Thus the top-hat
repulsion most strongly hinders binding among
the three realizations considered, reinforcing the conclusions in the main text. For an
attractive contact interaction ($b<0$),
the additional short-range attraction lowers
the critical Yukawa strength.

\subsection{Finite-range treatment of crossed pseudoscalar exchange}

In Sec.~\ref{sec:GP}, the $u$-channel exchange of the
pseudoscalar glueball was approximated by a contact
interaction evaluated at threshold. Here we examine
the effect of retaining its momentum dependence.
This provides a test of the local approximation
used in our main analysis.

The crossed contribution to the $GP\to GP$
amplitude is
\begin{equation}
\mc_u=-\frac{h^2}{u-m_P^2}
\text{ .}
\end{equation}
Unlike the $t$-channel exchange, the $u$-channel
involves a nonzero energy transfer at threshold,
$q^0=m_G-m_P$. Neglecting recoil corrections,
its nonrelativistic form reads
\begin{equation}
\mc_{u,\rm NR}(\boldsymbol p',\boldsymbol p)
\simeq
\frac{h^2}
{(\boldsymbol p'+\boldsymbol p)^2+\kappa_u^2}
\text{ , }
\kappa_u^2=m_G(2m_P-m_G)
\text{ .}
\end{equation}
The corresponding interaction is nonlocal in
the relative coordinate, acting as
\begin{equation}
[\hat V_u\psi](\boldsymbol r)
=-\alpha_u\frac{e^{-\kappa_u r}}{r}
\psi(-\boldsymbol r)
\text{ , }
\alpha_u=\frac{h^2}{16\pi m_Gm_P}
\text{ .}
\end{equation}
For an $S$ wave,
$\psi(-\boldsymbol r)=\psi(\boldsymbol r)$,
so the crossed exchange reduces to an additional
attractive Yukawa potential.

The resulting $S$-wave interaction is therefore
\begin{equation}\label{eq:GPcrossedpot}
V_{GP}^{\rm crossed}(r)
=-\alpha_{GP}\frac{e^{-m_Gr}}{r}
-\alpha_u\frac{e^{-\kappa_u r}}{r}
+(\beta_4+\beta_s)
\delta_{r_0}^{(3)}(\boldsymbol r)
\text{ .}
\end{equation}
The smeared repulsive contact contributions $\beta_4$
and $\beta_s$ are those defined in
Eq.~\eqref{eq:GPbeta}. Clearly, now the $u$-channel
contribution must not be included again in
the contact coefficient. Indeed, its
zero-momentum limit gives
\begin{equation}
\beta_u=-\frac{4\pi\alpha_u}{\kappa_u^2}
\text{ , }
\beta_{GP}=\beta_4+\beta_s+\beta_u
\text{ , }
\end{equation}
recovering the local matching used in the
main text.

For the central masses and $\Lambda_G=0.372$ GeV,
we obtain
\begin{equation}
\alpha_{GP}=6.293
\text{ , }
\alpha_u=5.642
\text{ , }
\kappa_u=2.415\ \mathrm{GeV}
\text{ , }
\beta_4+\beta_s=11.480\ \mathrm{GeV}^{-2}
\text{ .}
\end{equation}
The range of the crossed Yukawa interaction is
$\kappa_u^{-1}\simeq0.082$ fm, shorter than
the scalar Yukawa range $m_G^{-1}\simeq0.115$ fm.

We determine the first-binding threshold by
solving the zero-energy radial equation for
Eq.~\eqref{eq:GPcrossedpot}, using the same
normalized top-hat profile for the remaining
contact interaction. The results are compared
with the local approximation in
Table~\ref{tab:scan}.

The two prescriptions lead to comparable but different numerical
thresholds and, remarkably, to a qualitatively
different dependence on $r_0$. While the local
approximation exhibits only a weak, nearly
monotonic dependence, the explicit crossed-exchange
potential develops a minimum, similarly
to the $GG$ case. Among the radii listed in
Table~\ref{tab:scan}, the minimum occurs at
$r_0=0.10$ fm. A finer numerical scan places it
near $r_0\simeq0.082$ fm, where
$\Lambda_{G,\rm crit}^{GP}\simeq0.694$ GeV.
This nonmonotonic behavior reflects the competing
effects of the finite-range Yukawa attractions
and the remaining repulsive contact interaction,
whose influence eventually vanishes as its
spatial range shrinks.

The two prescriptions also differ in their formal
zero-radius limit. In the local approximation,
the attractive contact interaction leads to
$\Lambda_{G,\rm crit}^{GP}\propto r_0^{-1/2}$,
which diverges as $r_0\to0$. When the crossed
exchange is retained explicitly, however, the
remaining contact interaction is repulsive and
becomes ineffective in this limit. The potential
then approaches the sum of two attractive Yukawa
interactions, whose first-binding threshold is
$\Lambda_{G,\rm crit}^{GP}\simeq1.006$ GeV
for the central masses. These limiting results
characterize the two potential models and should
not be interpreted as predictions at arbitrarily
small physical interaction radii.

The central lattice-matched scale
$\Lambda_G=0.372$ GeV remains below all these
critical values. Thus, the prediction of $GP$
binding survives throughout the investigated
range of contact radii, although the quantitative
threshold is sensitive to the treatment of
crossed exchange.

\bibliographystyle{unsrtnat}
\bibliography{references}
\end{document}